\documentclass{appolb}
\usepackage{graphicx}

\usepackage{amsfonts}

\newcommand{\Pom}{\mathbb{P}}
\newcommand{\Ode}{\mathbb{O}}
\newcommand{\Reg}{\mathbb{R}}

\renewcommand\slash[1]{\not \! #1}

\newcommand{\bpta}{\mbox{\boldmath $p_{t,1}$}}
\newcommand{\bptb}{\mbox{\boldmath $p_{t,2}$}}

\newcommand{\bp}{\mbox{\boldmath $p$}}
\newcommand{\bk}{\mbox{\boldmath $k$}}

\usepackage{hyperref}

\begin{document}
\title{Searching for Odderon in Central Exclusive Processes \\at the LHC%
\thanks{Presented at XXVI Cracow EPIPHANY Conference, 
LHC Physics: Standard Model and Beyond, 7-10 January 2020}%
}
\author{Piotr Lebiedowicz
\address{Institute of Nuclear Physics Polish Academy of Sciences,\\
ul. Radzikowskiego 152, PL-31-342 Krak\'ow, Poland}
}
\maketitle
\begin{abstract}
There seem to be recently an evidence for $C = -1$
odderon exchange in proton-proton elastic scattering at high energies. 
Here we discuss three different central-exclusive-production processes
$pp \to pp (\phi \to K^{+}K^{-})$, 
$pp \to pp (\phi \to \mu^{+}\mu^{-})$, and
$pp \to pp \phi \phi \to pp K^{+}K^{-}K^{+}K^{-}$,
as a possible source of information for soft odderon exchange.
The theoretical results are calculated within the tensor-pomeron 
and vector-odderon model for soft reactions. 
We include absorptive corrections at the amplitude level.
To describe the low-energy data measured in the past
by the WA102 collaboration
we consider also subleading processes 
with reggeized vector-meson exchanges and reggeons. 
We discuss possible evidences for odderon exchange 
at the low energies and try to make predictions at the LHC.
\end{abstract}
  
\section{Introduction}

The odderon exchange became recently topical again.
So far there is no unambiguous experimental evidence for the odderon ($\Ode$),
the charge conjugation $C = -1$ counterpart of the $C= +1$ pomeron ($\Pom$),
introduced on theoretical grounds in \cite{Lukaszuk:1973nt, Joynson:1975az}.
The odderon was predicted in QCD as a colourless
$C$-odd three-gluon bound state exchange \cite{Kwiecinski:1980wb,Bartels:1980pe}.
A hint of the odderon was seen in ISR results~\cite{Breakstone:1985pe}
as a small difference between the differential cross sections
of elastic proton-proton ($pp$) and proton-antiproton ($p \bar{p}$) scattering
in the diffractive dip region at $\sqrt{s} = 53$~GeV.
The interpretation of this difference is, however, 
complicated due to non-negligible contributions from secondary reggeons. 
Recently the TOTEM Collaboration has published data from high-energy elastic
proton-proton scattering experiments at the LHC 
\cite{Antchev:2017yns,Antchev:2018rec}.
The interpretation of these results is controversial at the moment.
Some authors claim for instance that the $\rho$ measurements
show that there must be an odderon effect at $t = 0$ \cite{Martynov}. 
But other authors find that no odderon contribution is needed
at $t = 0$ \cite{Khoze,Broilo:2018qqs,Donnachie:2019ciz}.

As was discussed in \cite{Schafer:1991na} 
exclusive diffractive $J/\psi$ and $\phi$ production from the pomeron-odderon fusion
in high-energy $pp$ and $p\bar{p}$ collisions 
is a direct probe for a possible odderon exchange.
Exclusive production of heavy vector mesons,
$J/\psi$ and $\Upsilon$, from the pomeron-odderon 
and the pomeron-photon fusion
in the pQCD $k_{t}$-factorization approach was discussed in \cite{Bzdak:2007cz}.
However, so far in no one of the exclusive reactions 
a clear identification of the odderon was found experimentally.

A possible probe of the odderon is photoproduction of $C = +1$ mesons 
\cite{Schafer:1992pq,Barakhovsky:1991ra}.
At sufficiently high energies only odderon and photon exchange contribute to these reactions.
Photoproduction of the pseudoscalars $\pi^{0}$, $\eta$, $\eta'$, $\eta_{c}$, 
and of the tensor $f_{2}(1270)$ in $ep$ scattering at high energies 
was discussed in \cite{Kilian:1997ew,Berger:1999ca,Berger:2000wt,Donnachie:2005bu,Ewerz:2006gd}.
For a nice review of odderon physics see~\cite{Ewerz:2003xi}.
In \cite{Goncalves:2015hra} 
the measurement of the exclusive $\eta_{c}$ production
in nuclear collisions was discussed.
Recently, the possibility of probing the odderon 
in ultraperipheral proton-ion collisions was considered
\cite{Goncalves:2018pbr,Harland-Lang:2018ytk}.
The situation of the odderon in this context is also not obvious and requires further studies.

In \cite{Ewerz:2013kda} the tensor-pomeron and vector-odderon concept 
was introduced for soft reactions.
In this approach, the $C = +1$ pomeron and the reggeons 
$\Reg_{+} = f_{2 \Reg}, a_{2 \Reg}$ are treated as effective
rank-2 symmetric tensor exchanges
while the $C = -1$ odderon and the reggeons 
$\Reg_{-} = \omega_{\Reg}, \rho_{\Reg}$ are treated 
as effective vector exchanges.
Applications of the tensor-pomeron and vector-odderon ansatz
were given for photoproduction of pion pairs in \cite{Bolz:2014mya}
and for a number of central-exclusive-production (CEP) reactions 
in $pp$ collisions 
in \cite{CEP,
Lebiedowicz:2018sdt,
Lebiedowicz:2018eui,
Lebiedowicz:2019jru,
Lebiedowicz:2019boz}.
Also contributions from the subleading exchanges, $\Reg_{+}$ and $\Reg_{-}$, 
were discussed in these works.
As an example, for the $pp \to pp p\bar{p}$ reaction \cite{Lebiedowicz:2018sdt}
the contributions involving the odderon are expected to be small 
since its coupling to the proton is very small.
We have predicted asymmetries in the (pseudo)rapidity distributions 
of the centrally produced antiproton and proton.
The asymmetry is caused by interference effects of the dominant ($\Pom, \Pom$)
with the subdominant ($\Ode + \Reg_{-}$, $\Pom + \Reg_{+}$) 
and ($\Pom + \Reg_{+}$, $\Ode + \Reg_{-}$) exchanges.
We find for the odderon only very small effects,
roughly a factor 10 smaller than the effects due to reggeons.

In \cite{Ewerz:2016onn} the helicity structure of small-$|t|$ proton-proton elastic scattering
was considered in three models for the pomeron: tensor, vector, and scalar.
Only the tensor ansatz for the pomeron was found to be compatible with 
the high-energy experiment on polarized $pp$ elastic scattering \cite{Adamczyk:2012kn}.
In \cite{Britzger:2019lvc} the authors, 
using combinations of two tensor-type pomerons (a soft one and a hard one) 
and the $\Reg_{+}$-reggeon exchange,
successfully described low-$x$ deep-inelastic lepton-nucleon scattering and photoproduction.

In this talk we considered three exclusive processes
in proton-proton collisions
$pp \to pp (\phi \to K^{+} K^{-})$,
$pp \to pp (\phi \to \mu^{+} \mu^{-})$, and
$pp \to pp (\phi \phi \to K^{+} K^{-} K^{+} K^{-})$
in the nonperturbative tensor-pomeron and vector-odderon approach.
These processes were discussed in details in 
Refs.~\cite{Lebiedowicz:2019jru,Lebiedowicz:2019boz}.

\section{A sketch of formalism}

\begin{figure}
\centerline{
(a)\includegraphics[width=0.3\textwidth]{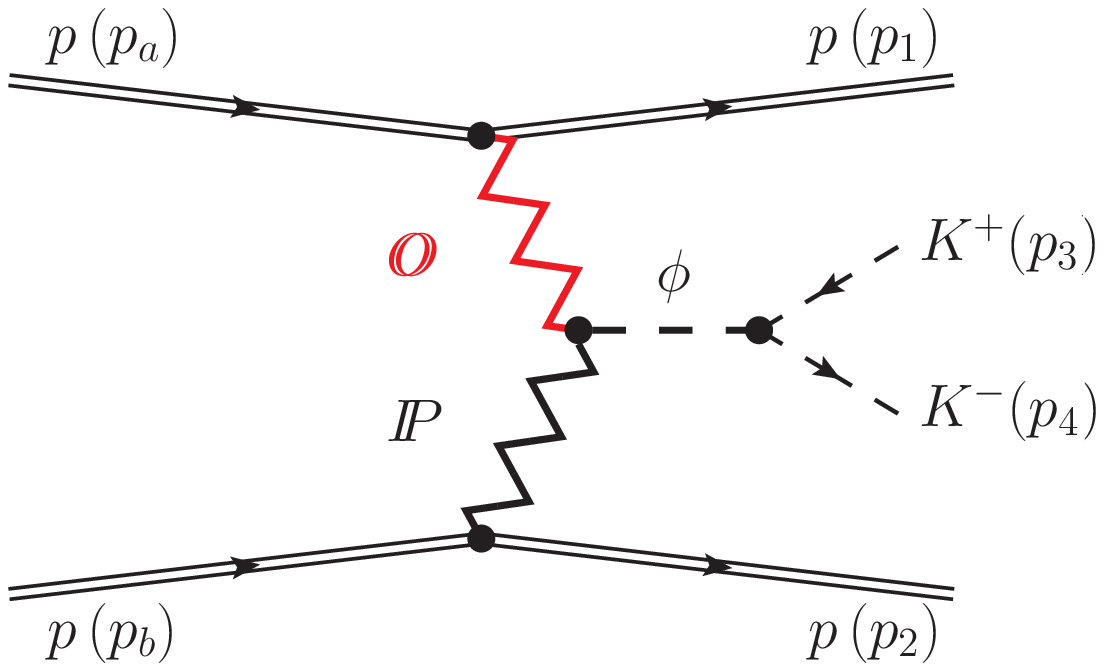} 
(b)\includegraphics[width=0.3\textwidth]{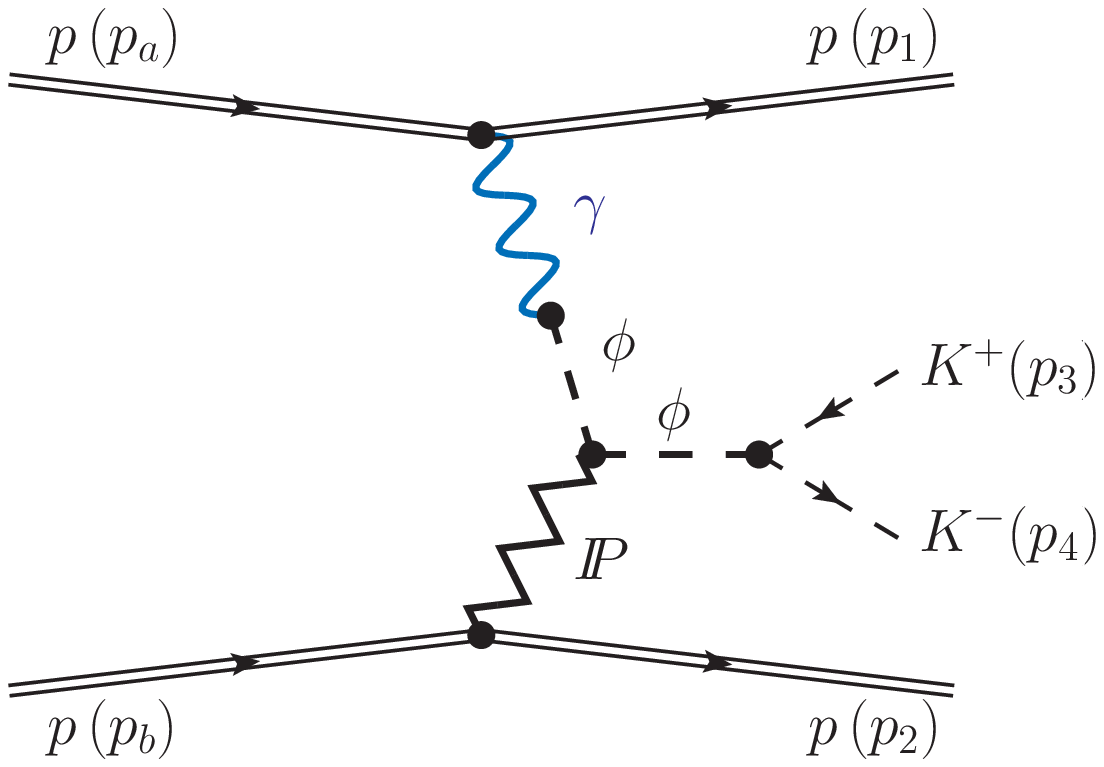} 
(c)\includegraphics[width=0.3\textwidth]{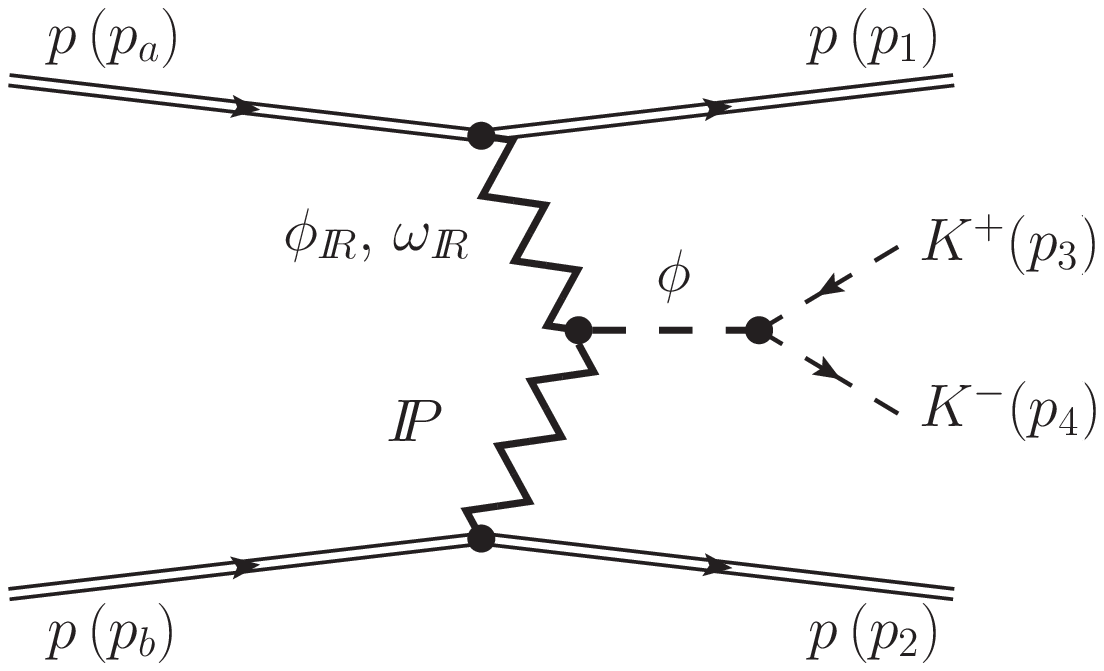}}
\caption{Some Born-level diagrams included in the analysis of single $\phi$ production:
(a)~odderon-pomeron fusion process;
(b)~photoproduction ($\gamma \Pom$ fusion) process;
(c)~reggeon-pomeron fusion processes. There are also the corresponding diagrams with the r{\^o}le of the protons interchanged, 
$( p\, (p_{a}), p\, (p_{1}) )
\leftrightarrow 
( p\, (p_{b}), p\, (p_{2}) )$.}
\label{fig:phi_diagrams}
\centerline{
(a)\includegraphics[width=0.29\textwidth]{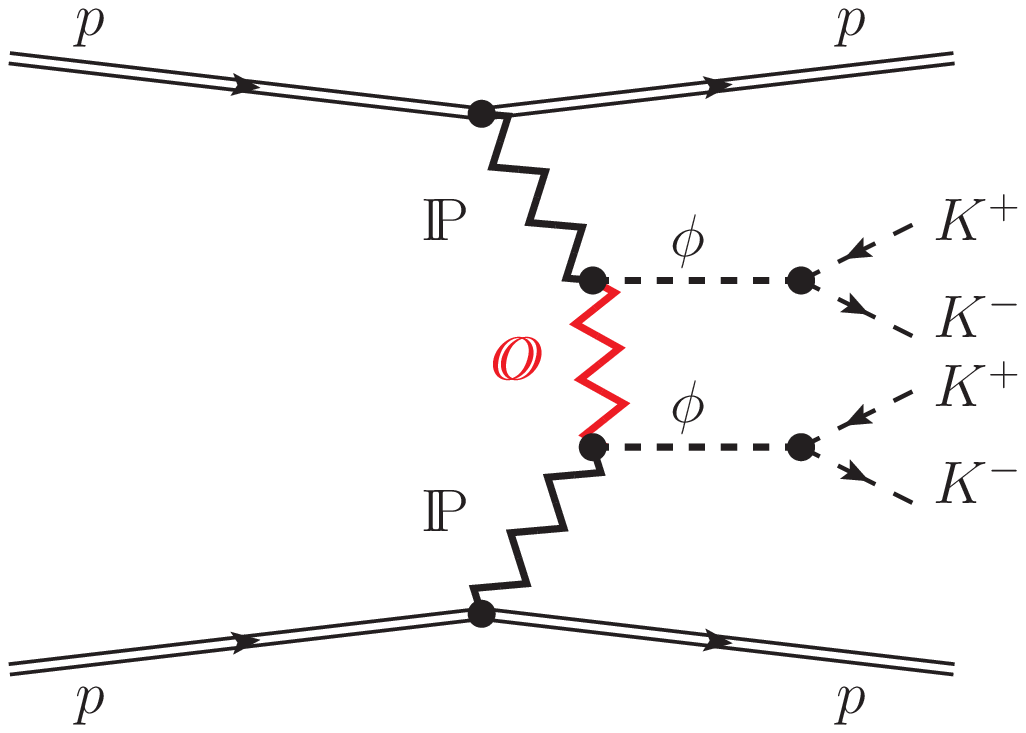}         
(b)\includegraphics[width=0.29\textwidth]{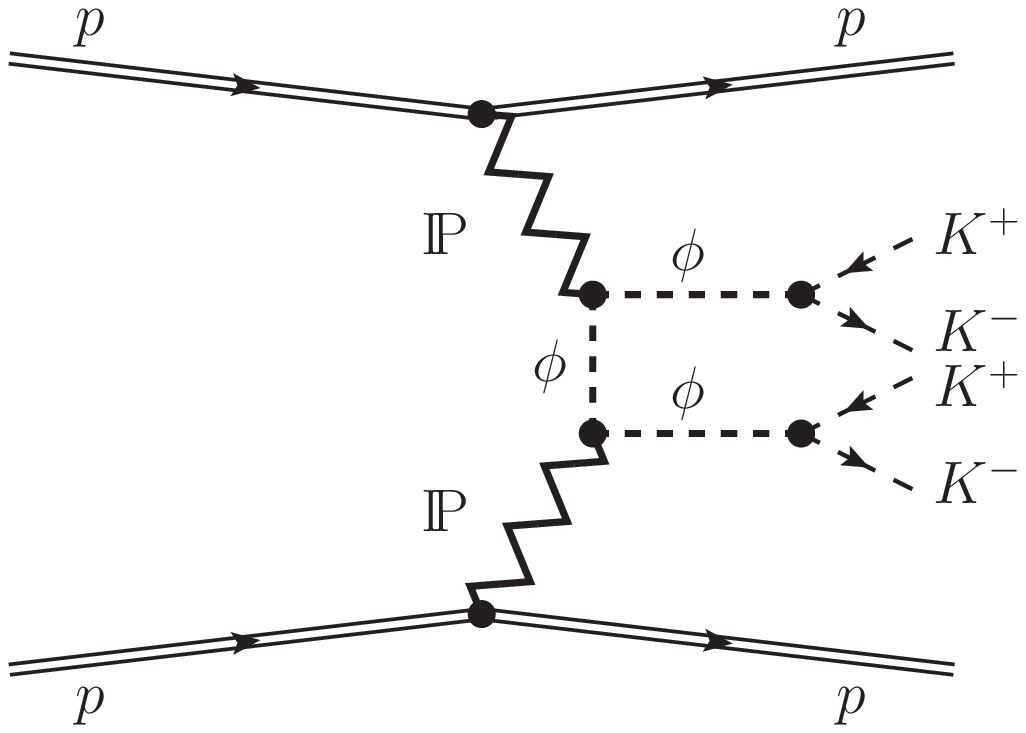}   
(c)\includegraphics[width=0.32\textwidth]{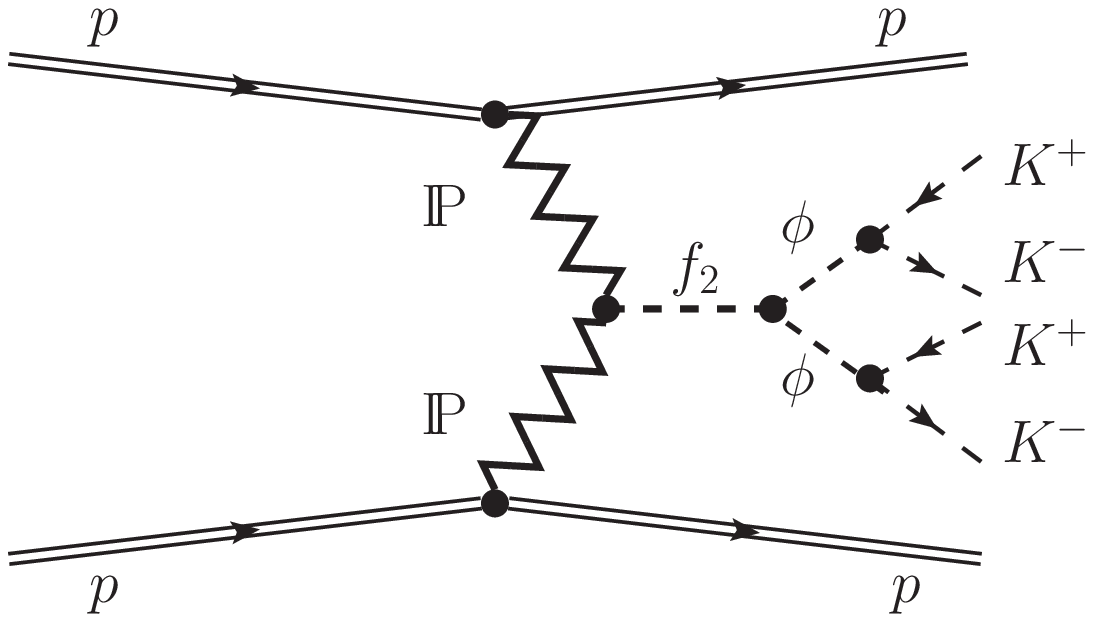}}
\caption{The Born-level diagrams 
for double pomeron central exclusive $\phi \phi$ production 
and their decays into $K^+ K^- K^+ K^-$:
(a)~continuum $\phi\phi$ production via an intermediate
odderon exchange;
(b) continuum via reggeized $\phi$-meson exchange;
(c) $\phi\phi$ production via an $f_{2}$ resonance.
Other resonances, e.g. of $f_{0}$- and $\eta$-type, can also contribute here; see Ref.~\cite{Lebiedowicz:2019jru}.}
\label{fig:4K_diagrams}
\end{figure}

For single $\phi$ production 
we include processes shown in Fig.~\ref{fig:phi_diagrams}.
For high energies and central $\phi \equiv \phi(1020)$ production we 
expect the reaction $pp \to pp \phi$
to be dominated by the fusion processes 
$\gamma \Pom \to \phi$ and $\Ode \Pom \to \phi$.
For the first process all couplings are, in essence, known.
For the odderon-exchange process we shall use the ans{\"a}tze 
from \cite{Ewerz:2013kda}
and we shall try to get information on the odderon parameters 
and couplings from the comparison to the WA102 data
for the $pp \to pp \phi$ and $pp \to pp \phi \phi$ reactions.
At the relatively low center-of-mass energy of the WA102 experiment, 
$\sqrt{s} = 29.1$~GeV, we have to include also 
subleading contributions with vector-meson 
(or reggeon) exchanges discussed in details 
in \cite{Lebiedowicz:2019boz}.

In the diffractive production of $\phi$ meson pairs, 
it is possible to have pomeron-pomeron fusion 
with intermediate $\hat{t}/\hat{u}$-channel odderon 
exchange \cite{Lebiedowicz:2019jru};
see the first diagram in Fig.~\ref{fig:4K_diagrams}.
Thus, the reaction
is a good candidate for the $\Ode$-exchange searches,
as it does not involve the coupling of the odderon to the proton
(the $\Ode$-$\Pom$-$\Ode$ contribution is negligibly small).

As an example, we consider the reaction
\begin{eqnarray}
p(p_{a}) + p(p_{b}) \to
p(p_{1}) + 
[ \phi(p_{34}) \to K^{+}(p_{3}) + K^{-}(p_{4}) ]
+ p(p_{2}) \,,
\label{2to4_reaction_KK_via_phi}
\end{eqnarray}
where $p_{a,b}$, $p_{1,2}$ 
denote the four-momenta of the protons and
$p_{3,4}$ denote the four-momenta of the $K$ mesons, respectively.
The kinematic variables are
\begin{eqnarray}
&&p_{34} = p_{3} + p_{4}\,, \quad q_1 = p_{a} - p_{1}\,, \quad q_2 = p_{b} - p_{2}\,,
\nonumber \\ 
&&s = (p_{a} + p_{b})^{2} = (p_{1} + p_{2} + p_{34})^{2}\,,
\nonumber \\ 
&&t_1 = q_{1}^{2}\,, 
\quad t_2 = q_{2}^{2}\,,
\quad s_{1} = (p_{1} + p_{34})^{2}\,, 
\quad s_{2} = (p_{2} + p_{34})^{2}\,.
\label{2to4_kinematic}
\end{eqnarray}

The Born-level amplitude for the diffractive production of the $\phi(1020)$ 
via odderon-pomeron fusion,
see diagram~(a) in Fig.~\ref{fig:phi_diagrams},
can be written as
\begin{eqnarray}
&&{\cal M}^{(\Ode \Pom)}_{pp \to pp K^{+}K^{-}} 
= (-i)
\bar{u}(p_{1}, \lambda_{1}) 
i\Gamma^{(\Ode pp)}_{\mu}(p_{1},p_{a}) 
u(p_{a}, \lambda_{a}) \nonumber \\
&& \qquad \times  
i\Delta^{(\Ode)\,\mu \rho_{1}}(s_{1}, t_{1}) \,
i\Gamma^{(\Pom \Ode \phi)}_{\rho_{1} \rho_{2} \alpha \beta}(-q_{1},p_{34})\, 
i\Delta^{(\phi)\,\rho_{2} \kappa}(p_{34})\,
i\Gamma^{(\phi KK)}_{\kappa}(p_{3},p_{4})
\nonumber \\
&&\qquad \times 
i\Delta^{(\Pom)\,\alpha \beta, \delta \eta}(s_{2},t_{2}) \,
\bar{u}(p_{2}, \lambda_{2}) 
i\Gamma^{(\Pom pp)}_{\delta \eta}(p_{2},p_{b}) 
u(p_{b}, \lambda_{b}) \,.
\label{amplitude_oderon_pomeron}
\end{eqnarray}
%
The effective propagator and the proton vertex function
for tensorial pomeron are as follows \cite{Ewerz:2013kda}:
\begin{eqnarray}
&&i \Delta^{(\Pom)}_{\mu \nu, \kappa \lambda}(s,t) =
\frac{1}{4s} \left( g_{\mu \kappa} g_{\nu \lambda} 
                  + g_{\mu \lambda} g_{\nu \kappa}
                  - \frac{1}{2} g_{\mu \nu} g_{\kappa \lambda} \right)
(-i s \alpha'_{\Pom})^{\alpha_{\Pom}(t)-1}\,,
\label{A1}\\
&&i\Gamma_{\mu \nu}^{(\Pom pp)}(p',p)
=-i 3 \beta_{\Pom NN} F_{1}( (p'-p)^{2} ) \nonumber \\
&& \qquad \qquad \qquad \; \times 
\left\lbrace 
\frac{1}{2} 
\left[ \gamma_{\mu}(p'+p)_{\nu} 
     + \gamma_{\nu}(p'+p)_{\mu} \right]
- \frac{1}{4} g_{\mu \nu} ( \slash{p}' + \slash{p} )
\right\rbrace\,,
\label{A4} 
\end{eqnarray}
where $\beta_{\Pom NN} = 1.87$~GeV$^{-1}$ and $t = (p'-p)^{2}$.
For simplicity we use the electromagnetic Dirac form factor $F_{1}(t)$ of the proton.
The pomeron trajectory $\alpha_{\Pom}(t)$
is assumed to be of standard linear form 
(see, e.g., \cite{Donnachie:2002en}):
$\alpha_{\Pom}(t) = \alpha_{\Pom}(0)+\alpha'_{\Pom}\,t$,
$\alpha_{\Pom}(0) = 1.0808$,
$\alpha'_{\Pom} = 0.25 \; {\rm GeV}^{-2}$.

Our ansatz for the $C = -1$ odderon follows (3.16), (3.17) and (3.68), (3.69) 
of \cite{Ewerz:2013kda}:
\begin{eqnarray}
&&i \Delta^{(\Ode)}_{\mu \nu}(s,t) = 
-i g_{\mu \nu} \frac{\eta_{\Ode}}{M_{0}^{2}} \,(-i s \alpha'_{\Ode})^{\alpha_{\Ode}(t)-1}\,,
\label{A12} \\
&&i\Gamma_{\mu}^{(\Ode pp)}(p',p) 
= -i 3\beta_{\Ode pp} \,M_{0}\,F_{1} ( (p'-p)^{2} ) \,\gamma_{\mu}\,,
\label{A13}
\end{eqnarray}
where $\eta_{\Ode}$ is a parameter with value $\eta_{\Ode} = \pm 1$;
$M_{0} = 1$~GeV is inserted for dimensional reasons.
We assumed for the odderon trajectory
$\alpha_{\Ode}(t) = \alpha_{\Ode}(0)+\alpha'_{\Ode}\,t.$
%
In our calculations we shall choose as default values 
$\alpha_{\Ode}(0) = 1.05$,
$\alpha'_{\Ode} = 0.25 \,\mathrm{GeV}^{-2}$, 
and $\eta_{\Ode} = - 1$; see \cite{Lebiedowicz:2019boz}.
We assumed 
%
$\beta_{\Ode pp} = 0.1 \,\beta_{\Pom NN}$.
%

For the $\Pom \Ode \phi$ vertex we use an ansatz analogous to the $\Pom \phi \phi$ vertex;
see (3.48)--(3.50) of \cite{Lebiedowicz:2019jru}.
We get then with $(-q_{1},\rho_{1})$ and $(p_{34},\rho_{2})$
the outgoing oriented momenta and the vector indices
of the odderon and the $\phi$ meson, respectively,
and $\alpha \beta$ the pomeron indices,
\begin{eqnarray}
i\Gamma^{(\Pom \Ode \phi)}_{\rho_{1} \rho_{2} \alpha \beta}(-q_{1},p_{34}) 
&=&
i \left[ 2\,a_{\Pom \Ode \phi}\, \Gamma^{(0)}_{\rho_{2} \rho_{1} \alpha \beta}(p_{34},-q_{1})
- b_{\Pom \Ode \phi}\,\Gamma^{(2)}_{\rho_{2} \rho_{1} \alpha \beta}(p_{34},-q_{1}) \right] \nonumber\\
&&\times F_{M}(q_{2}^{2})\,F_{M}(q_{1}^{2})\,F^{(\phi)}(p_{34}^{2}) \,.
\label{A15}
\end{eqnarray}  
Here we use the relations (3.20) of \cite{Ewerz:2013kda}
and as in (3.49) of \cite{Lebiedowicz:2019jru}
we take the factorised form for the $\Pom \Ode \phi$ form factor; see \cite{Lebiedowicz:2019boz}.
The coupling parameters $a_{\Pom \Ode \phi}$, $b_{\Pom \Ode \phi}$ in (\ref{A15})
and the cut-off parameter $\Lambda_{0,\,\Pom \Ode \phi}^{2}$ 
in $F_{M}(t) = 1/(1 - t/\Lambda_{0,\,\Pom \Ode \phi}^{2})$
could be adjusted to experimental data.
The WA102 data allow us to determine the respective
coupling constants as $a_{\Pom \Ode \phi} = -0.8$~GeV$^{-3}$, $b_{\Pom \Ode \phi}= 1.6$~GeV$^{-1}$, and
$\Lambda_{0,\,\Pom \Ode \phi}^{2} = 0.5$~GeV$^{2}$;
see Sec. IV~A of \cite{Lebiedowicz:2019boz}.
We have checked that these parameters 
are compatible with our analysis of the WA102 data for
the $pp \to pp \phi \phi$ reaction in \cite{Lebiedowicz:2019jru}.

The full form of the vector-meson propagator is given by (3.2) of \cite{Ewerz:2013kda}.
Here we take the simple Breit-Wigner expression
as discussed in \cite{Lebiedowicz:2018eui}.
For the $\phi KK$ vertex we follow (4.24)--(4.26) of 
\cite{Lebiedowicz:2018eui}. For the details see Ref.~\cite{Lebiedowicz:2019boz}.

To give the full physical amplitude
we should include absorptive corrections to the Born amplitudes;
see e.g. \cite{Lebiedowicz:2015eka}.
The full amplitude includes the $pp$-rescattering corrections 
in the eikonal approximation is written as
\begin{eqnarray}
&&{\cal {M}} =
{\cal {M}}^{\mathrm{Born}} + 
{\cal {M}}^{\mathrm{abs.}}\,,\\
%
&&{\cal M}^{\mathrm{abs.}}(s,\bp_{1t},\bp_{2t})= 
\frac{i}{8 \pi^{2} s} \int d^{2} \bk_{t} \,
{\cal M}^{\mathrm{Born}}(s,\tilde{\bp}_{1t},\tilde{\bp}_{2t})\,
{\cal M}_{\mathrm{el}}^{(\Pom)}(s,-{\bk}_{t}^{2})\,, \quad
\label{abs_correction}
\end{eqnarray}
where $\tilde{\bp}_{1t} = {\bp}_{1t} - {\bk}_{t}$ and
$\tilde{\bp}_{2t} = {\bp}_{2t} + {\bk}_{t}$.
${\cal M}_{\mathrm{el}}^{(\Pom)}$ 
is the elastic $pp$-scattering amplitude
with the momentum transfer $t=-{\bk}_{t}^{2}$.

\section{Results}

\begin{figure}\centerline{
\includegraphics[width=0.44\textwidth]{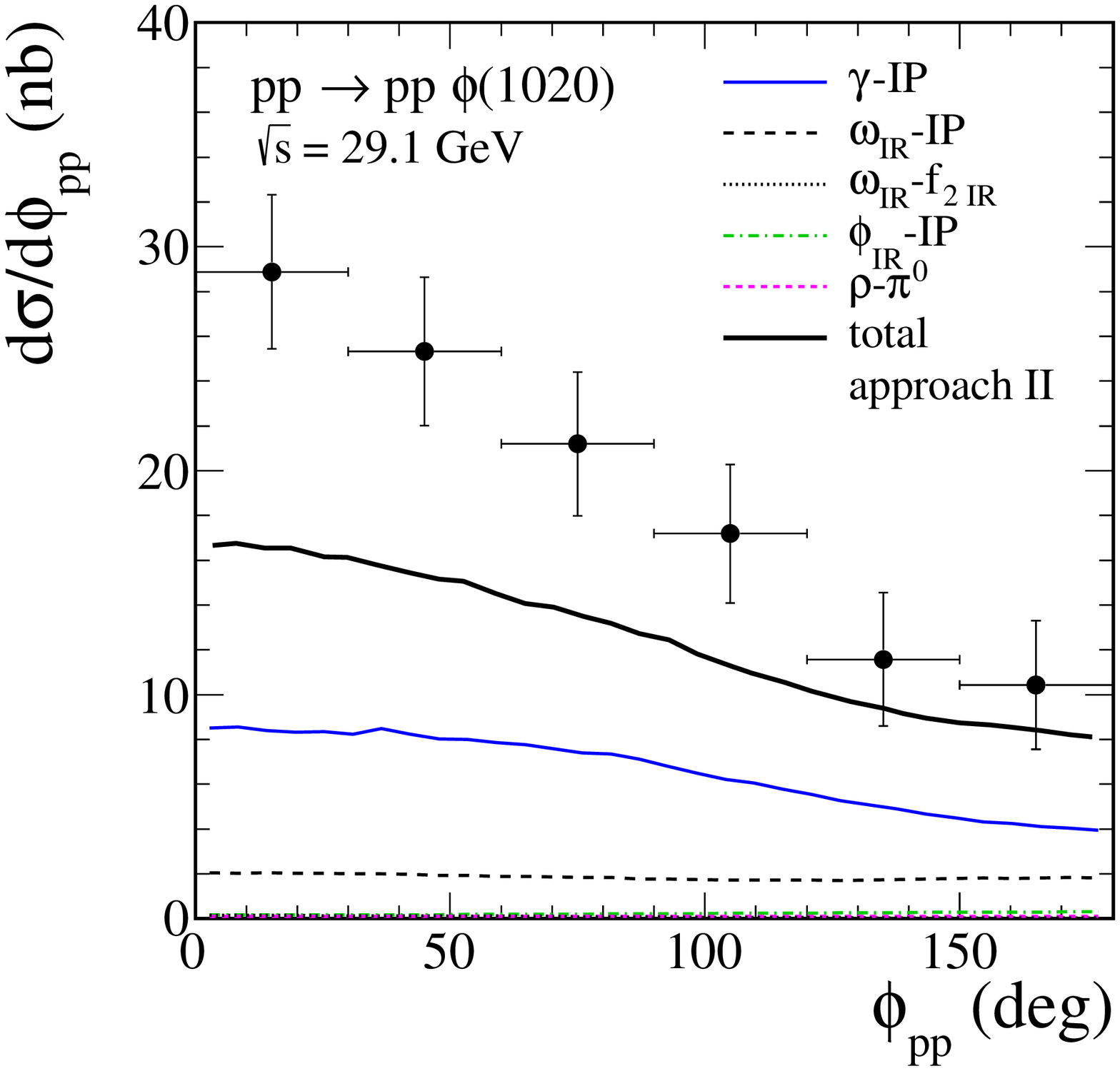}
\includegraphics[width=0.44\textwidth]{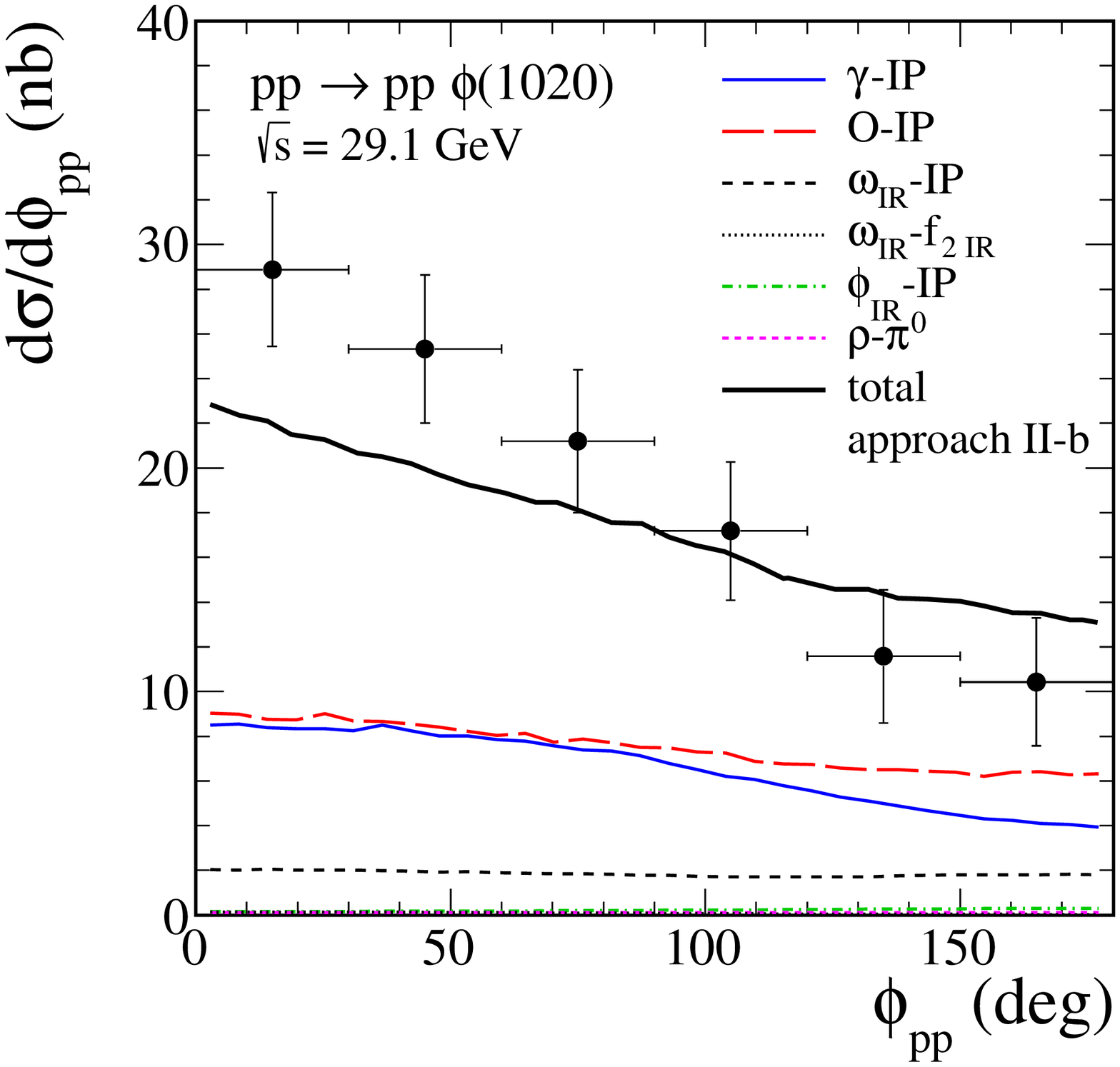}}
\caption{The distributions in azimuthal angle $\phi_{pp}$ 
between the transverse momentum vectors 
$\bpta$, $\bptb$ of the outgoing protons
for $\sqrt{s} = 29.1$~GeV 
together with the WA102 experimental data points 
normalized to the central value 
of the total cross section $\sigma_{\rm exp} = 60$~nb from \cite{Kirk:2000ws}.
In the left panel the results for the fusion processes 
$\gamma$-$\Pom$, $\omega_{\Reg}$-$\Pom$, $\omega_{\Reg}$-$f_{2 \Reg}$, $\phi_{\Reg}$-$\Pom$,
and $\rho$-$\pi^{0}$ are presented.
The coherent sum of all terms is shown by the black solid line.
In the right panel we added the $\Ode$-$\Pom$ fusion term
(see the red long-dashed line).}
\label{fig:WA102_phi}
\end{figure}

It is very difficult to describe the WA102 data from 
\cite{Kirk:2000ws}
for the $pp \to pp \phi$ reaction including the
$\gamma \Pom$-fusion mechanism only.
The result of our analysis is shown 
in Fig.~\ref{fig:WA102_phi}.
Inclusion of the odderon-exchange contribution significantly
improves the description of the $pp$ azimuthal correlations 
and the ${\rm dP_{t}}$ ``glueball-filter variable'' dependence 
of $\phi$ CEP measured by WA102; 
see the discussion in \cite{Lebiedowicz:2019boz}.
The absorption effects are included here 
and in the calculations presented below. 
To describe the low-energy data more accurately we consider also subleading fusion processes. 
Here we present results for the approach~II of \cite{Lebiedowicz:2019boz}.
We can see that the complete results indicate 
a large interference effect between the
$\gamma$-$\Pom$, $\Ode$-$\Pom$, $\omega_{\Reg}$-$\Pom$, $\omega_{\Reg}$-$f_{2 \Reg}$, $\phi_{\Reg}$-$\Pom$,
and $\rho$-$\pi^{0}$ (for the reggeized $\rho^{0}$-meson exchange) terms.
However, the subleading terms do not play a significant role at the LHC.

Having fixed the parameters of our model to the WA102 data we wish
to show our predictions at $\sqrt{s} = 13$~TeV for the LHC.
Here we focus on the limited dikaon invariant mass region 
i.e., the $\phi(1020)$ resonance region,
$1.01 \, {\rm GeV} < M_{K^{+}K^{-}} < 1.03\, {\rm GeV}$.

\begin{figure}[!ht]
\begin{center}
\includegraphics[width=0.44\textwidth]{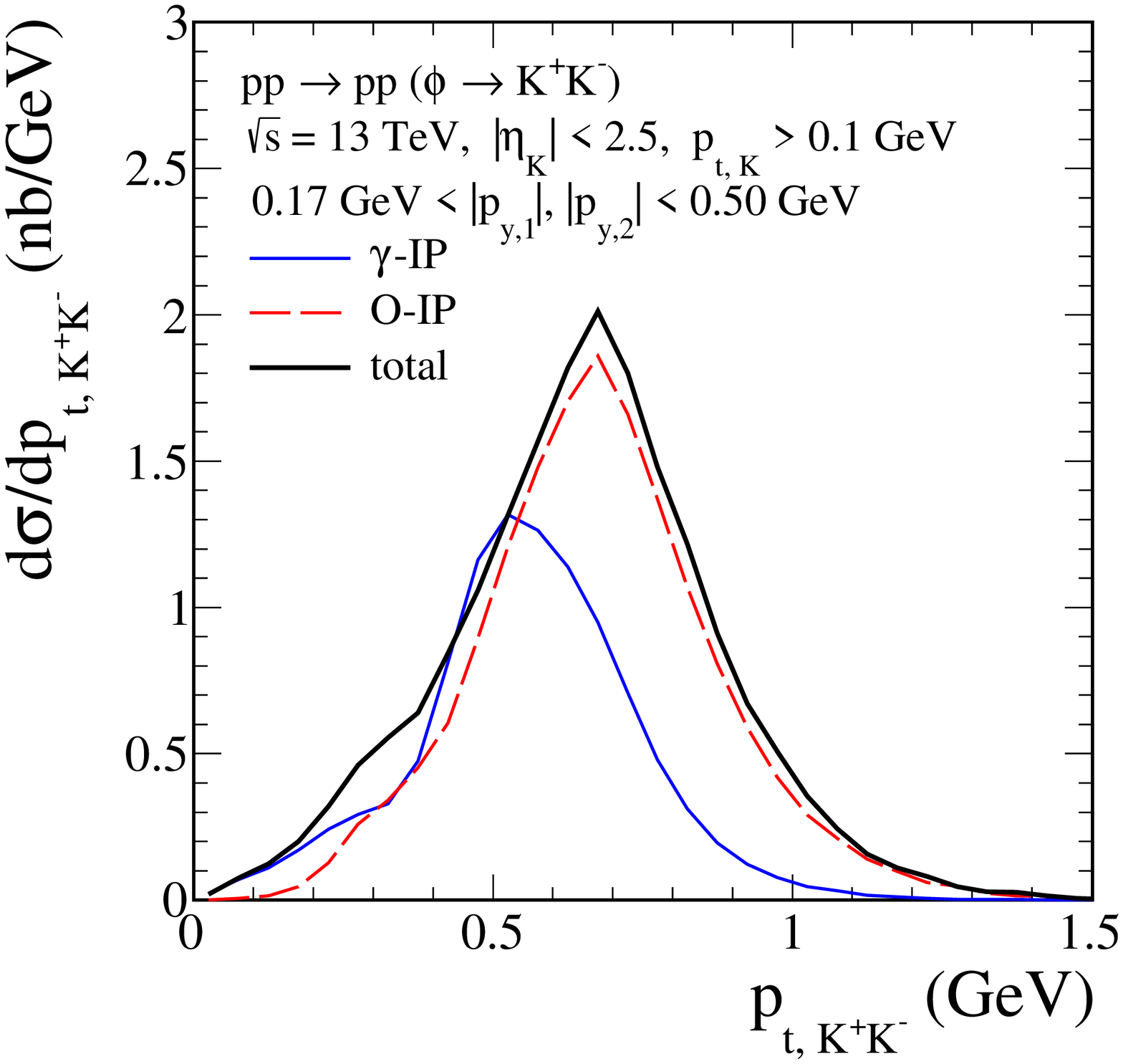}
\includegraphics[width=0.44\textwidth]{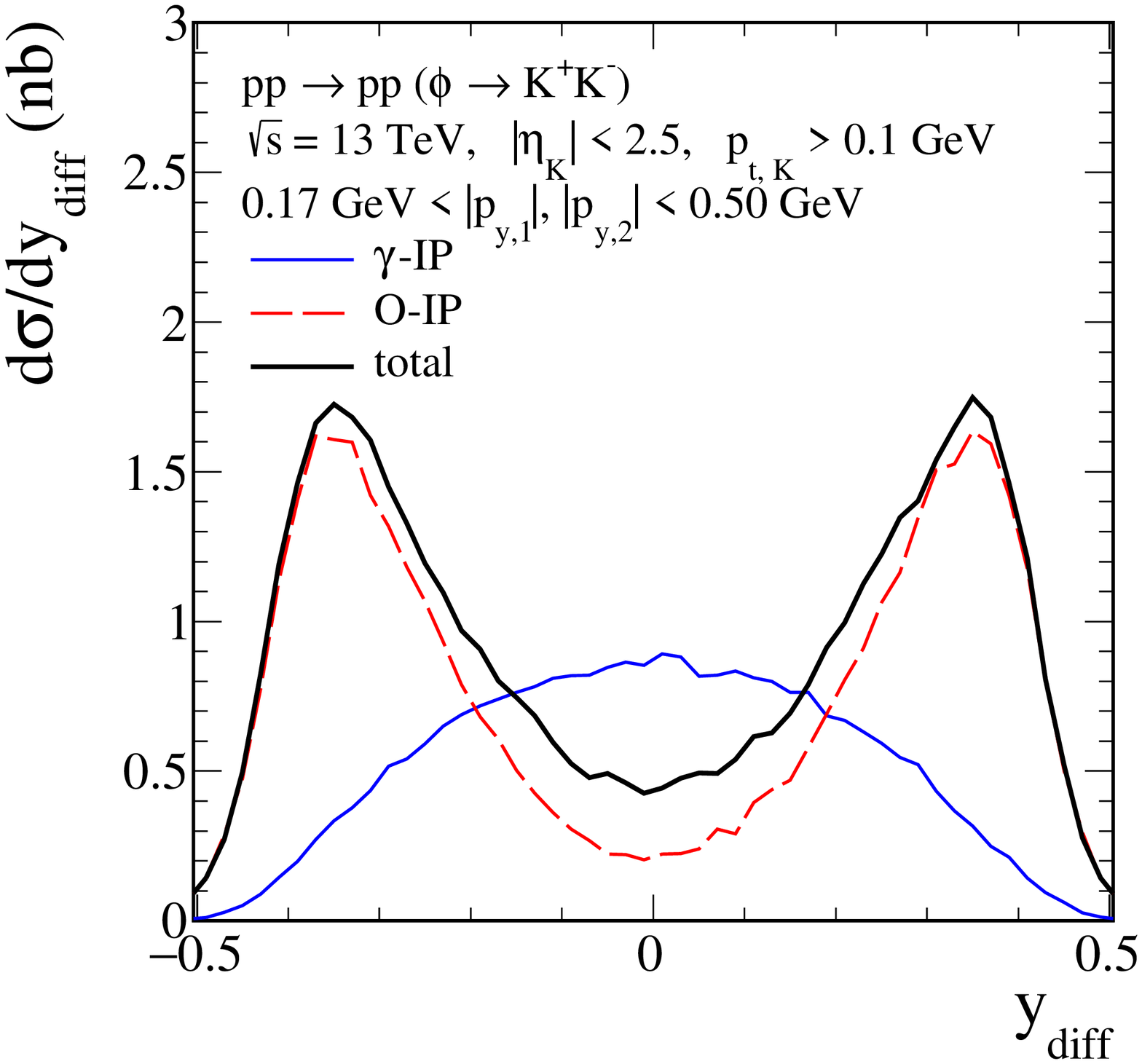}
\includegraphics[width=0.44\textwidth]{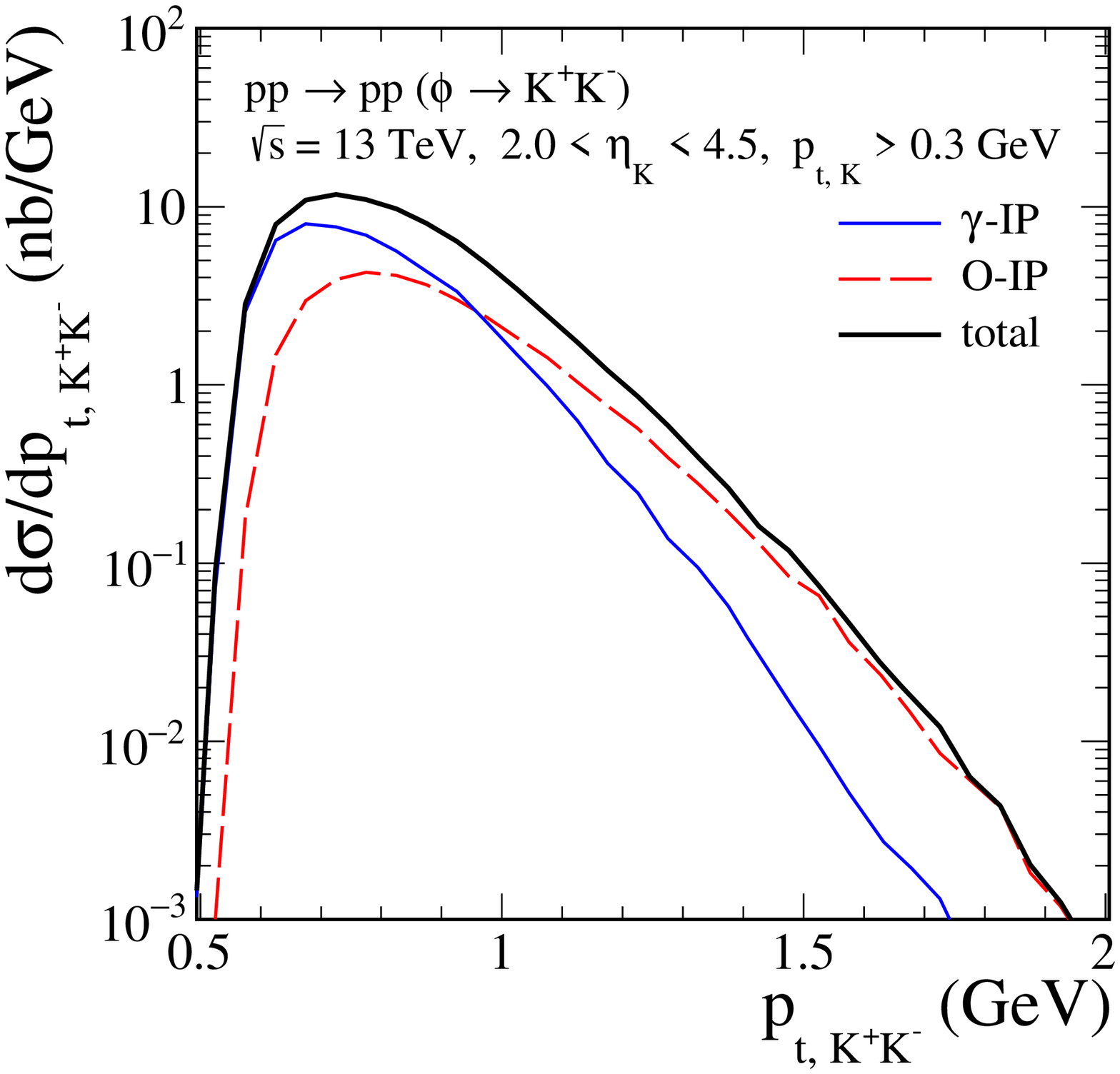}
\includegraphics[width=0.44\textwidth]{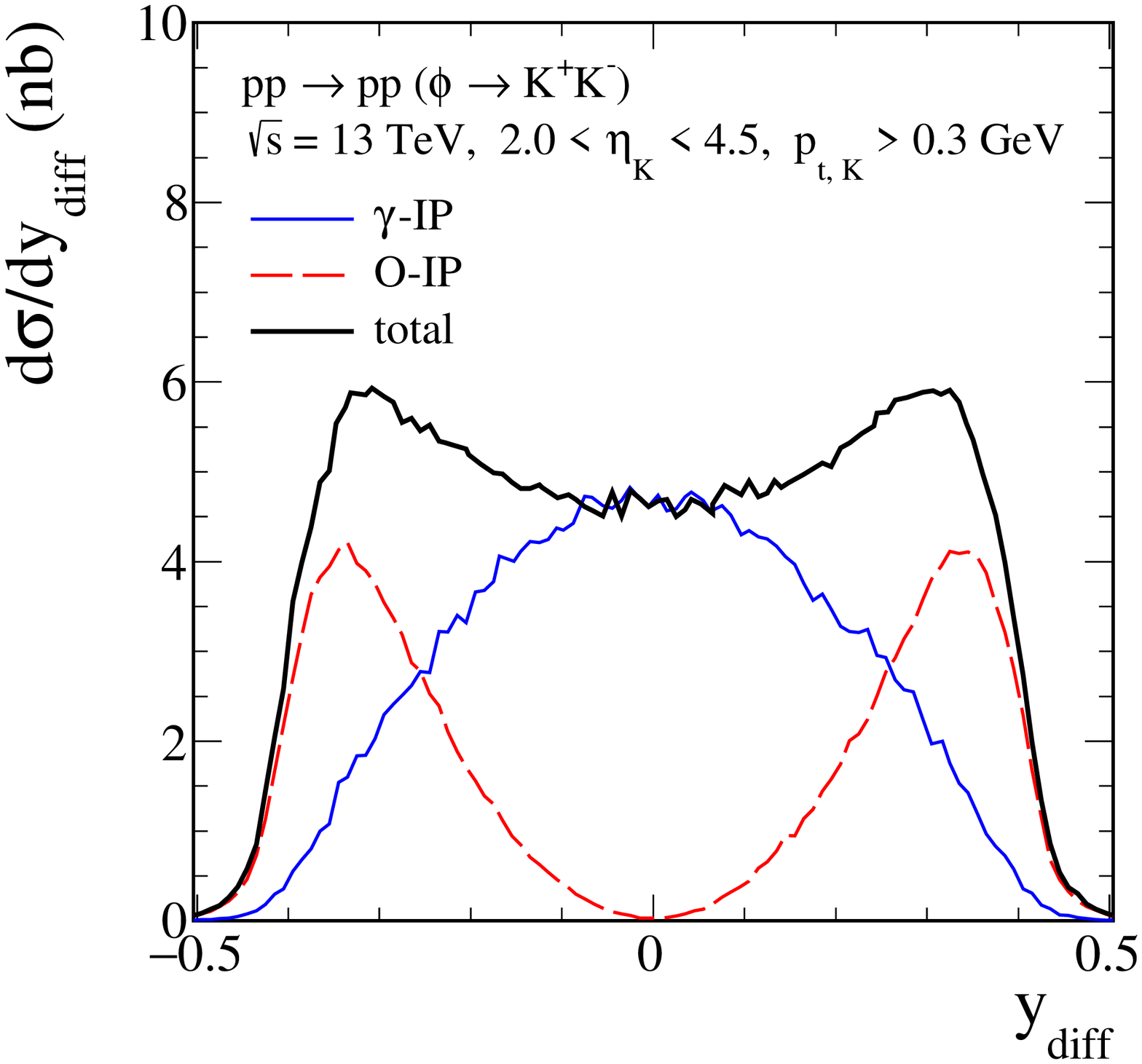}
\caption{\label{fig:_LHC_KK}
The distributions in transverse momentum of the $K^{+}K^{-}$ pair
(left) and in rapidity difference between kaons $\rm{y_{diff}}$
for the $pp \to pp (\phi \to K^{+}K^{-})$ reaction.
Shown are the $\gamma$-$\Pom$- and $\Ode$-$\Pom$-fusion contributions and their coherent sum (denoted by ``total'')
for the ATLAS-ALFA (top) and the LHCb (bottom) 
experimental cuts.}
\end{center}
\end{figure}

In Fig.~\ref{fig:_LHC_KK} 
we show the results for the $pp \to pp (\phi \to K^{+}K^{-})$
reaction for experimental conditions relevant for ATLAS-ALFA or CMS-TOTEM
($|\eta_{K}| < 2.5$, $p_{t, K} > 0.1$~GeV, 
and with extra cuts on the leading protons of 
0.17~GeV~$< |p_{y,1}|, |p_{y,2}|<$~0.50~GeV
as will be the proton momentum window for the ALFA detectors)
and at forward rapidities and without measurement of protons
relevant for LHCb.
The odderon-pomeron contribution dominates at larger
$p_{t, K^{+}K^{-}}$ and
$|\rm{y_{diff}}|$ compared to the photon-pomeron contribution.
For larger kaon transverse momenta 
(or transverse momentum of the $K^+ K^-$ pair)
the odderon-exchange contribution
is bigger than the photon-exchange one;
see Table~II of \cite{Lebiedowicz:2019boz}.
For the ATLAS-ALFA kinematics the absorption effects lead to a large
damping of the cross sections both for the hadronic diffractive 
and for the photoproduction mechanisms.

Now we discuss the $pp \to pp \mu^{+}\mu^{-}$ reaction
for the LHCb kinematics ($2.0 < \eta_{\mu} < 4.5$, $p_{t, \mu} > 0.1$~GeV).
Here we require no detection of the leading protons.
In contrast to dikaon production here there is 
for both the $\gamma$-$\Pom$- and the $\Ode$-$\Pom$-fusion contributions
a maximum at $\rm{y_{diff}} = 0$.
Imposing a larger cuts on the transverse momenta of the muons
reduces the continuum ($\gamma \gamma \to \mu^{+}\mu^{-}$) contribution which, however,
still remains sizeable at $\rm{y_{diff}} = 0$.
The dimuon-continuum process ($\gamma \gamma \to \mu^{+}\mu^{-}$)
was discussed e.g. in \cite{Lebiedowicz:2018muq}
in the context of the ATLAS measurement.

\begin{figure}[!ht]
\begin{center}
\includegraphics[width=0.44\textwidth]{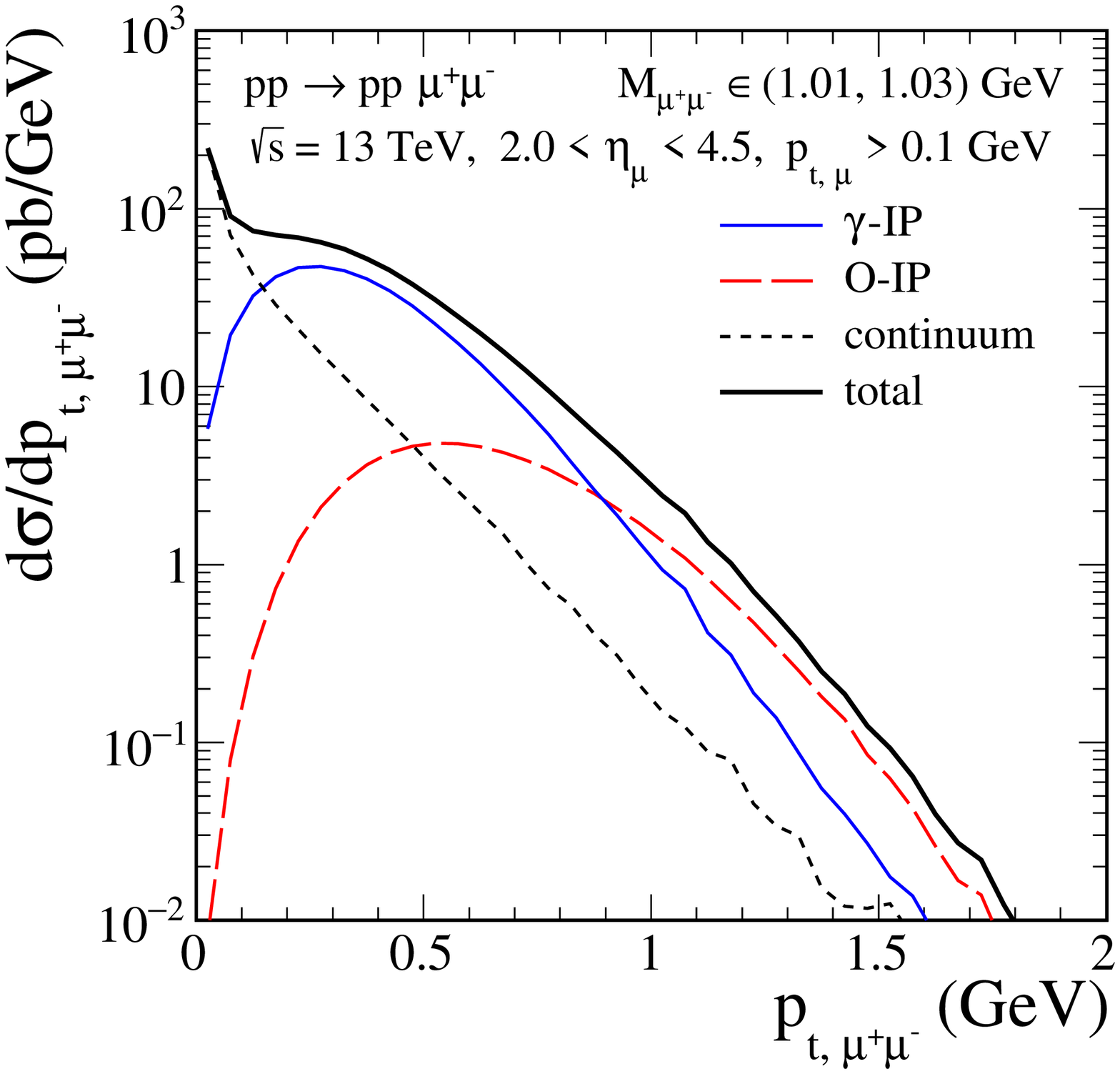}
\includegraphics[width=0.44\textwidth]{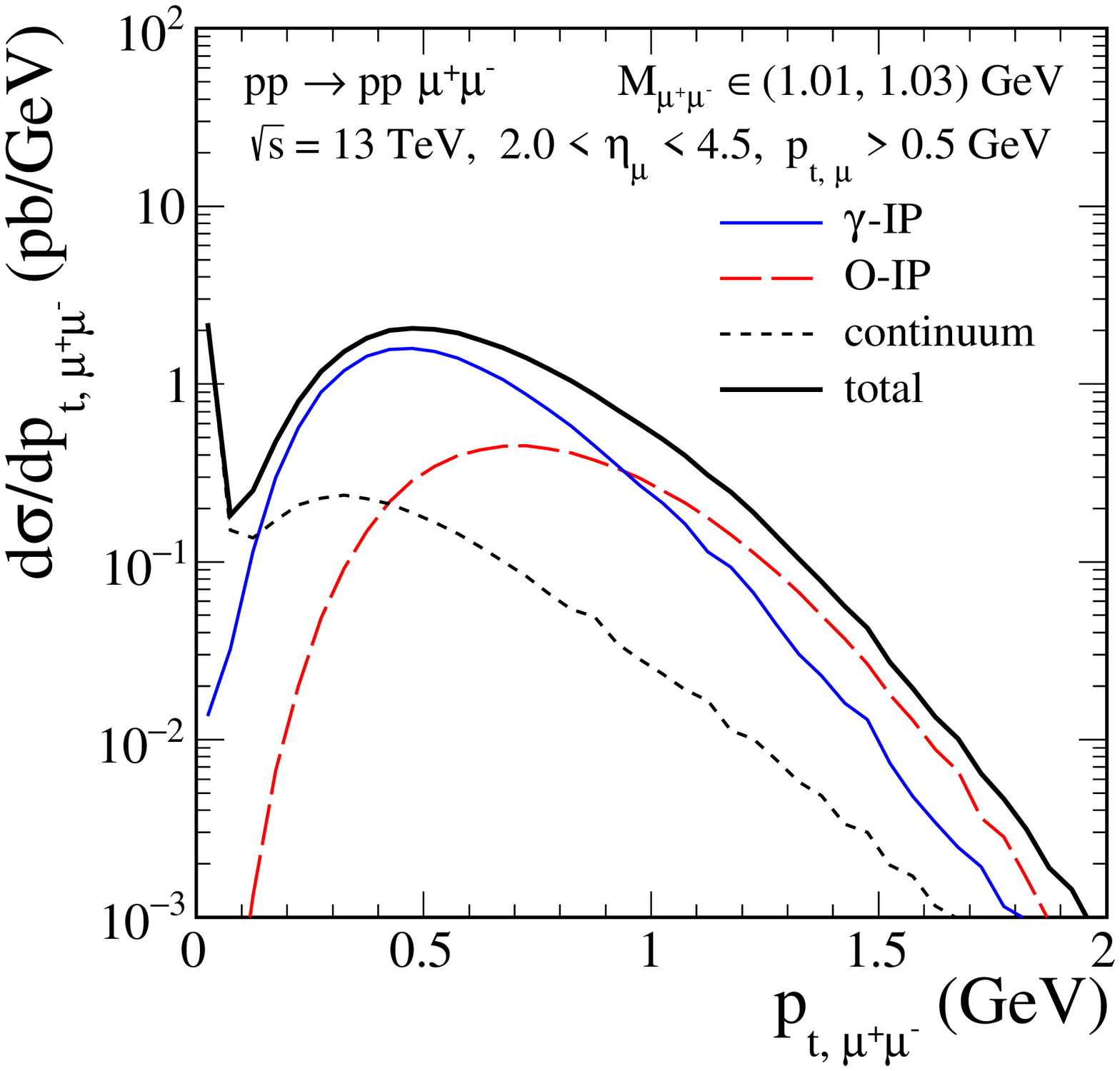}
\caption{\label{fig:LHCb_mumu_3}
The distributions in transverse momentum of the $\mu^{+}\mu^{-}$ pair
for the $pp \to pp \mu^{+}\mu^{-}$ reaction
in the dimuon invariant mass region $M_{\mu^{+}\mu^{-}} \in (1.01, 1.03)$~GeV.
Calculations were done for $\sqrt{s} = 13$~TeV, $2.0 < \eta_{\mu} < 4.5$
and for $p_{t, \mu} > 0.1$~GeV (left)
and for $p_{t, \mu} > 0.5$~GeV (right).
Results for the $\phi$-meson production via
the $\gamma$-$\Pom$- and the $\Ode$-$\Pom$-fusion processes
and the nonresonant $\gamma \gamma \to \mu^{+}\mu^{-}$ continuum term
are shown.
Their coherent sum is shown by the black solid line.}
\end{center}
\end{figure}

\begin{figure}[!ht]
\begin{center}
\includegraphics[width=0.44\textwidth]{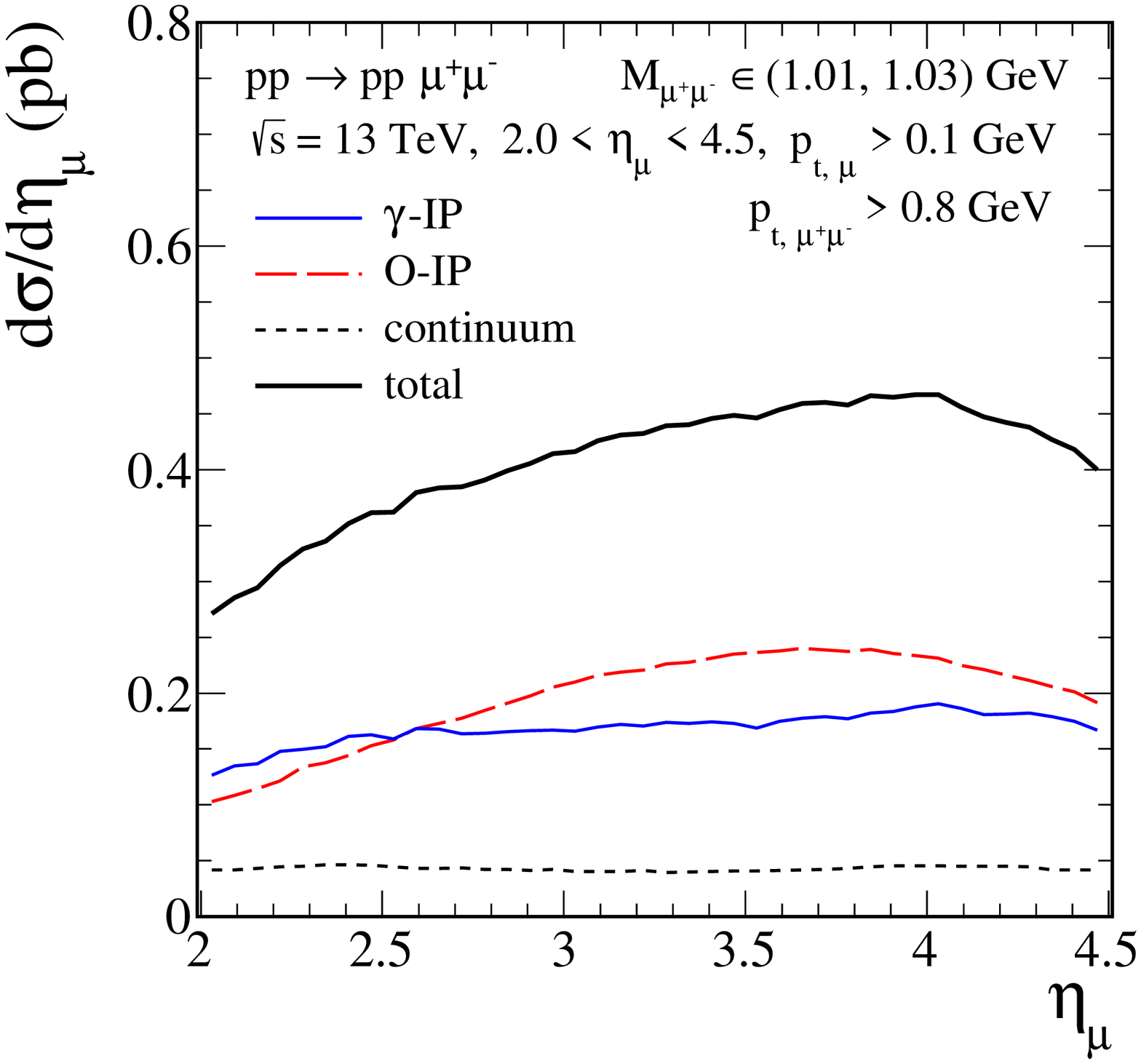}
\includegraphics[width=0.44\textwidth]{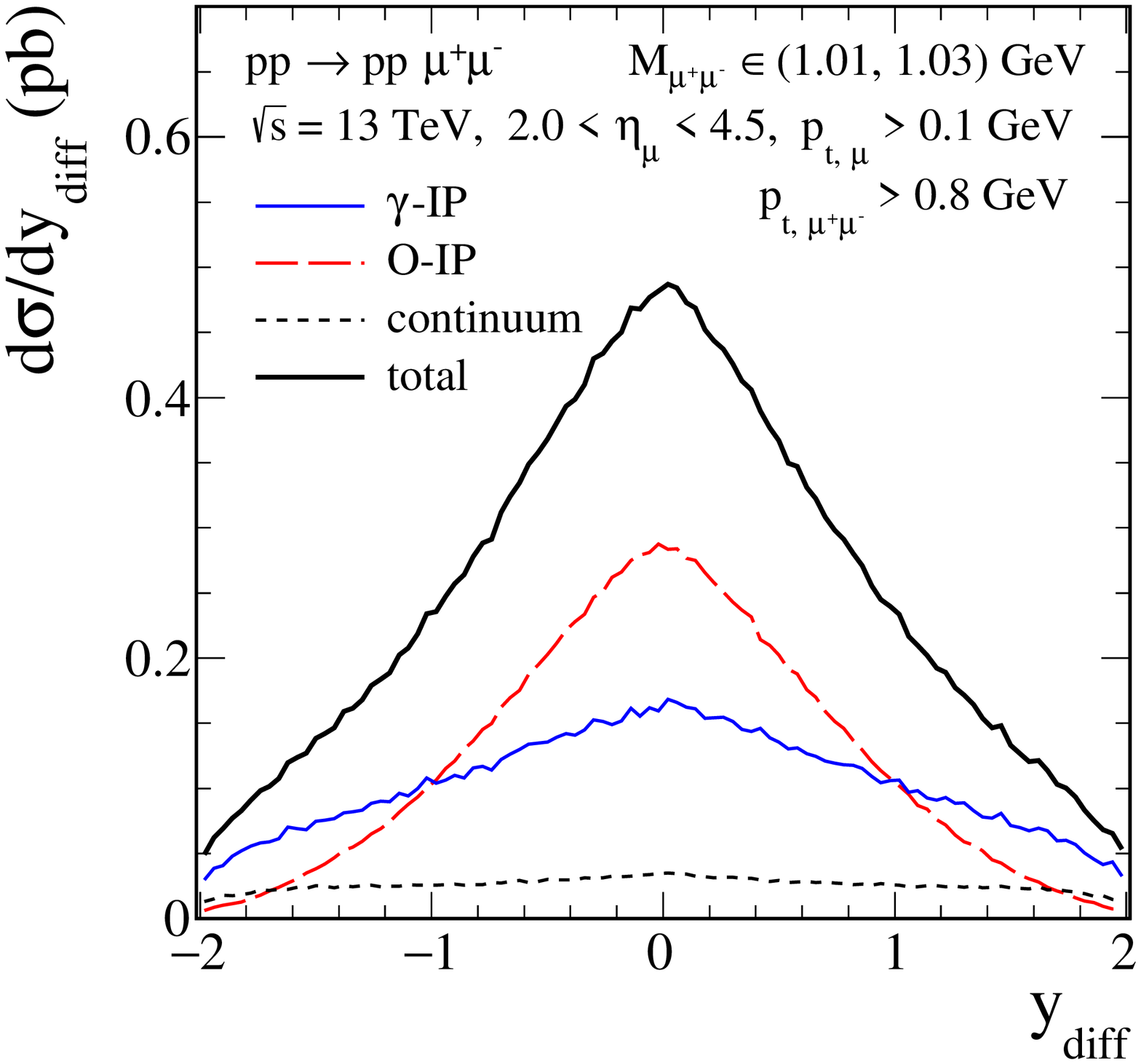}
\caption{\label{fig:LHCb_mumu_4}
The differential cross sections for the $pp \to pp \mu^{+}\mu^{-}$ reaction
in the dimuon invariant mass region $M_{\mu^{+}\mu^{-}} \in (1.01, 1.03)$~GeV.
Calculations were done for $\sqrt{s} = 13$~TeV, $2.0 < \eta_{\mu} < 4.5$, 
$p_{t, \mu} > 0.1$~GeV, and $p_{t, \mu^{+}\mu^{-}} > 0.8$~GeV.
The meaning of the lines is the same as 
in Fig.~\ref{fig:LHCb_mumu_3}.}
\end{center}
\end{figure}

Figure~\ref{fig:LHCb_mumu_3} shows the distributions 
in transverse momentum of the $\mu^{+}\mu^{-}$ pair.
We can see that the low-$p_{t,\mu^{+}\mu^{-}}$ cut can be helpful
to reduce the continuum and $\gamma$-$\Pom$-fusion contributions.
In Fig.~\ref{fig:LHCb_mumu_4} we show the results
when imposing in addition a cut $p_{t, \mu^{+}\mu^{-}} > 0.8$~GeV.
The $\gamma \gamma \to \mu^{+}\mu^{-}$ contribution is now very small.
We can see from the $\rm{y_{diff}}$ distribution that the photon-pomeron term gives
a broader distribution than the odderon-pomeron term. 
At $\rm{y_{diff}} = 0$ the odderon-exchange term 
is now bigger than the photoproduction terms.

\begin{figure}[!ht]
\begin{center}
\includegraphics[width=0.44\textwidth]{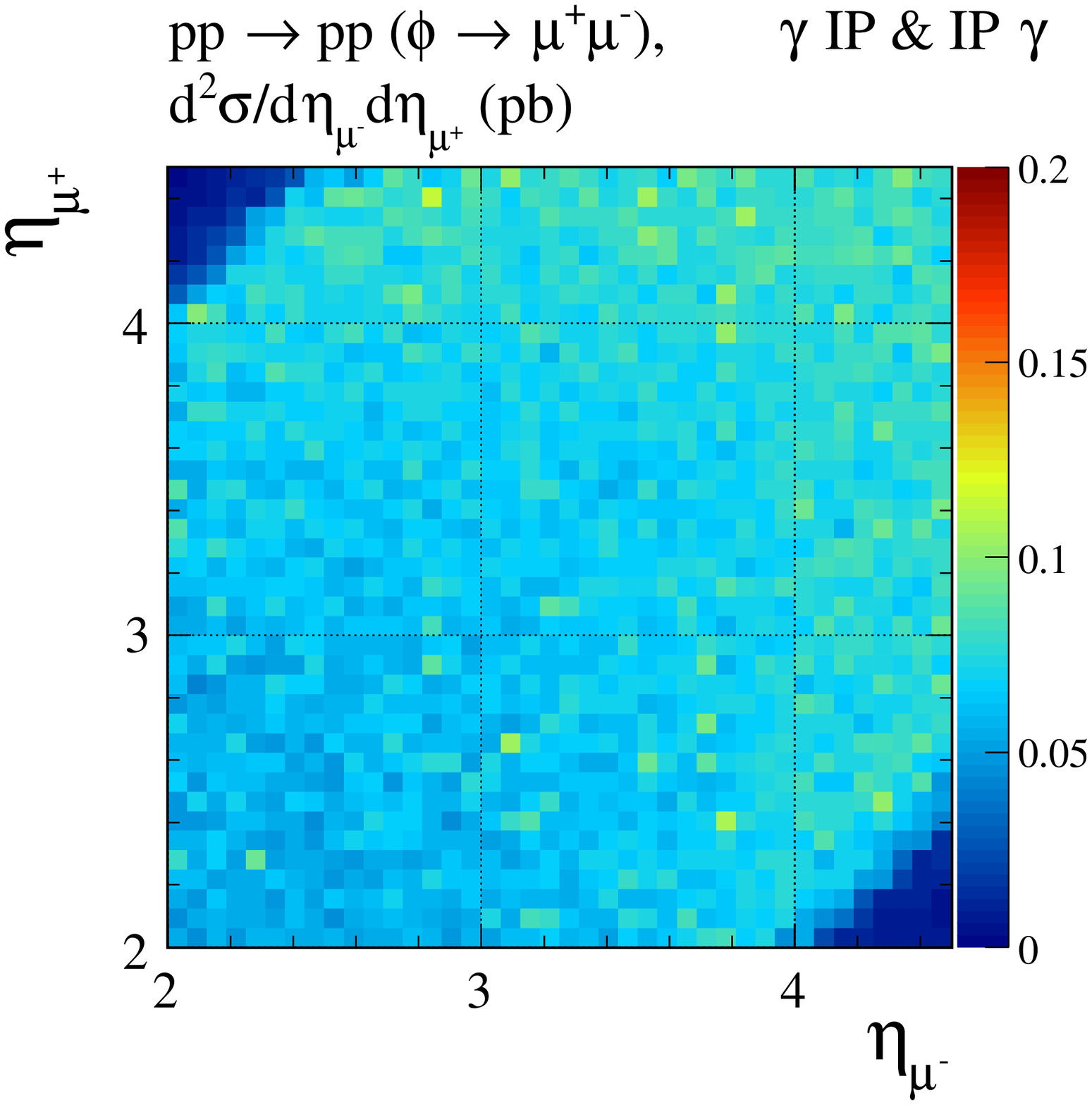}
\includegraphics[width=0.44\textwidth]{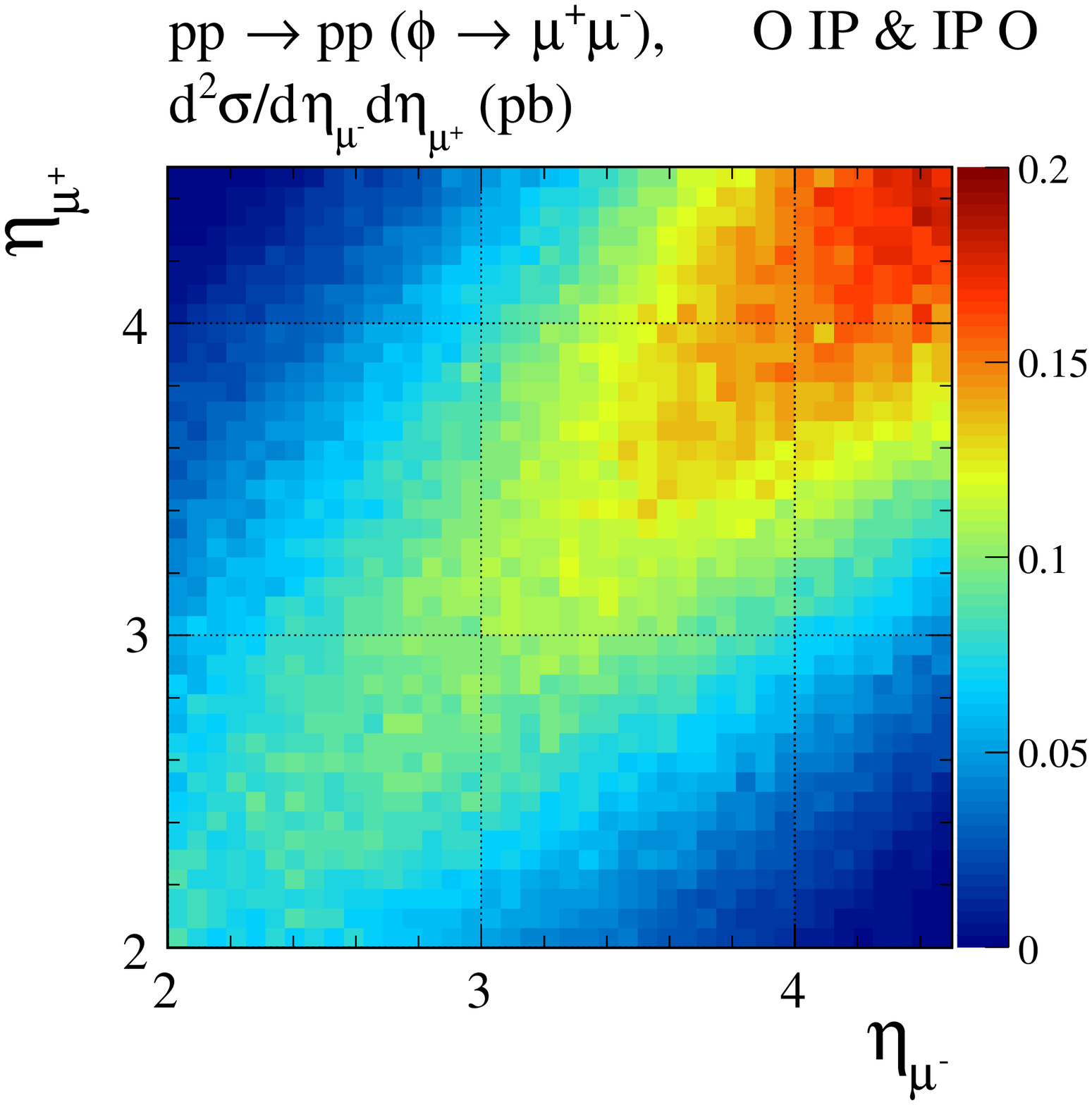}
\caption{\label{fig:LHCb_mumu_2D_eta3eta4}
\small
The two-dimensional distributions in ($\eta_{\mu^{+}}$, $\eta_{\mu^{-}}$)
for the $pp \to pp \mu^{+}\mu^{-}$ reaction.
Shown are the results for the $\phi$ production 
via $\gamma$-$\Pom$ fusion (left) and
via $\Ode$-$\Pom$ fusion (right). 
The calculations were done for $\sqrt{s} = 13$~TeV and with cuts on 
$M_{\mu^{+}\mu^{-}} \in (1.01, 1.03)$~GeV, $2.0 < \eta_{\mu} < 4.5$,
$p_{t, \mu} > 0.1$~GeV, and $p_{t, \mu^{+}\mu^{-}} > 0.8$~GeV.}
\end{center}
\end{figure}

In Fig.~\ref{fig:LHCb_mumu_2D_eta3eta4} we show
two-dimensional distributions in ($\eta_{\mu^{+}}$, $\eta_{\mu^{-}}$).
One can see quite different distributions
for the $\gamma$-$\Pom$- and the $\Ode$-$\Pom$ contributions. 
The odderon-exchange contribution shows an enhancement 
at $\eta_{\mu^{-}} \sim \eta_{\mu^{+}} \sim 4$.

\begin{figure}
\begin{center}
\includegraphics[width=0.44\textwidth]{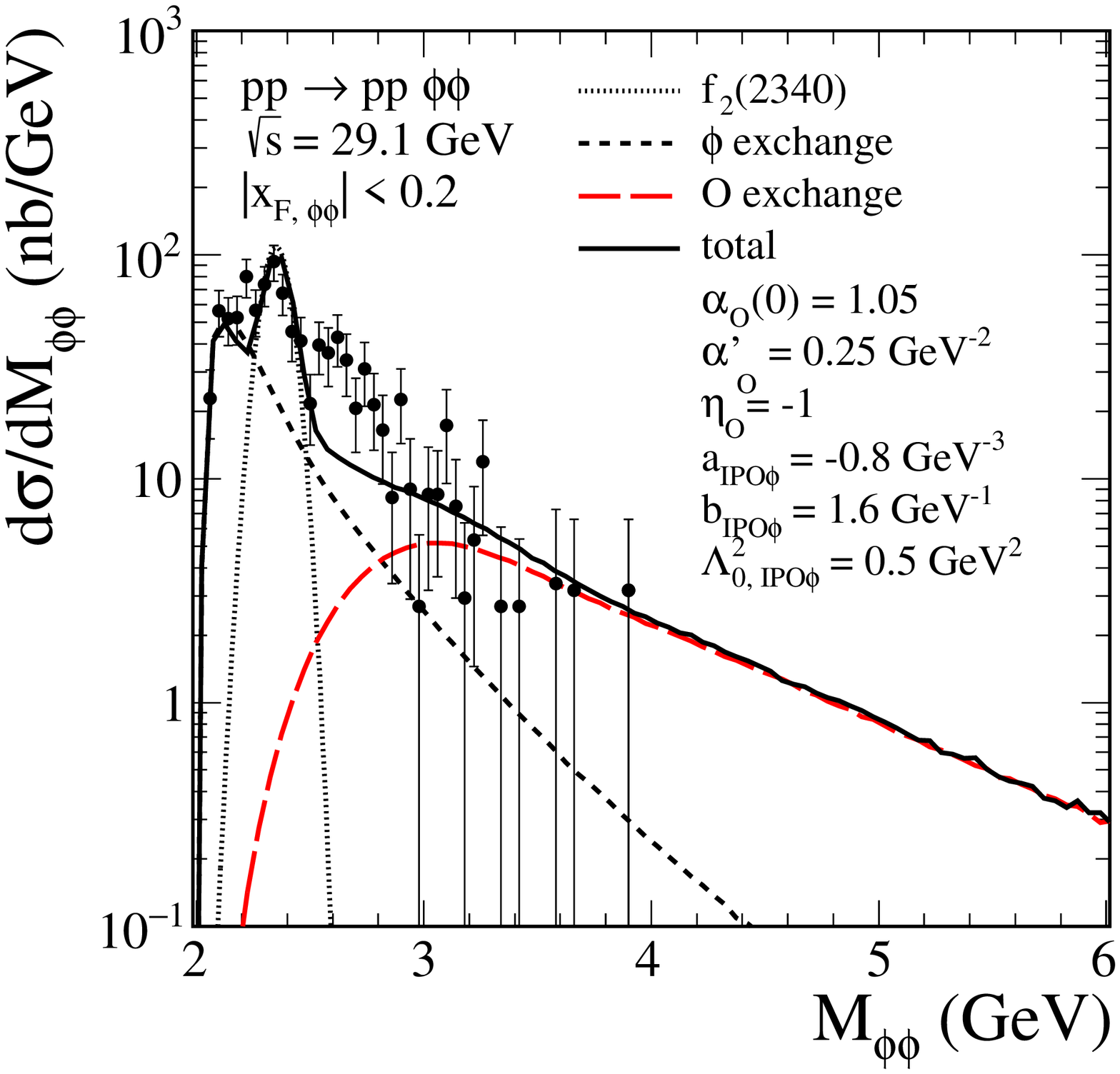}
\includegraphics[width=0.44\textwidth]{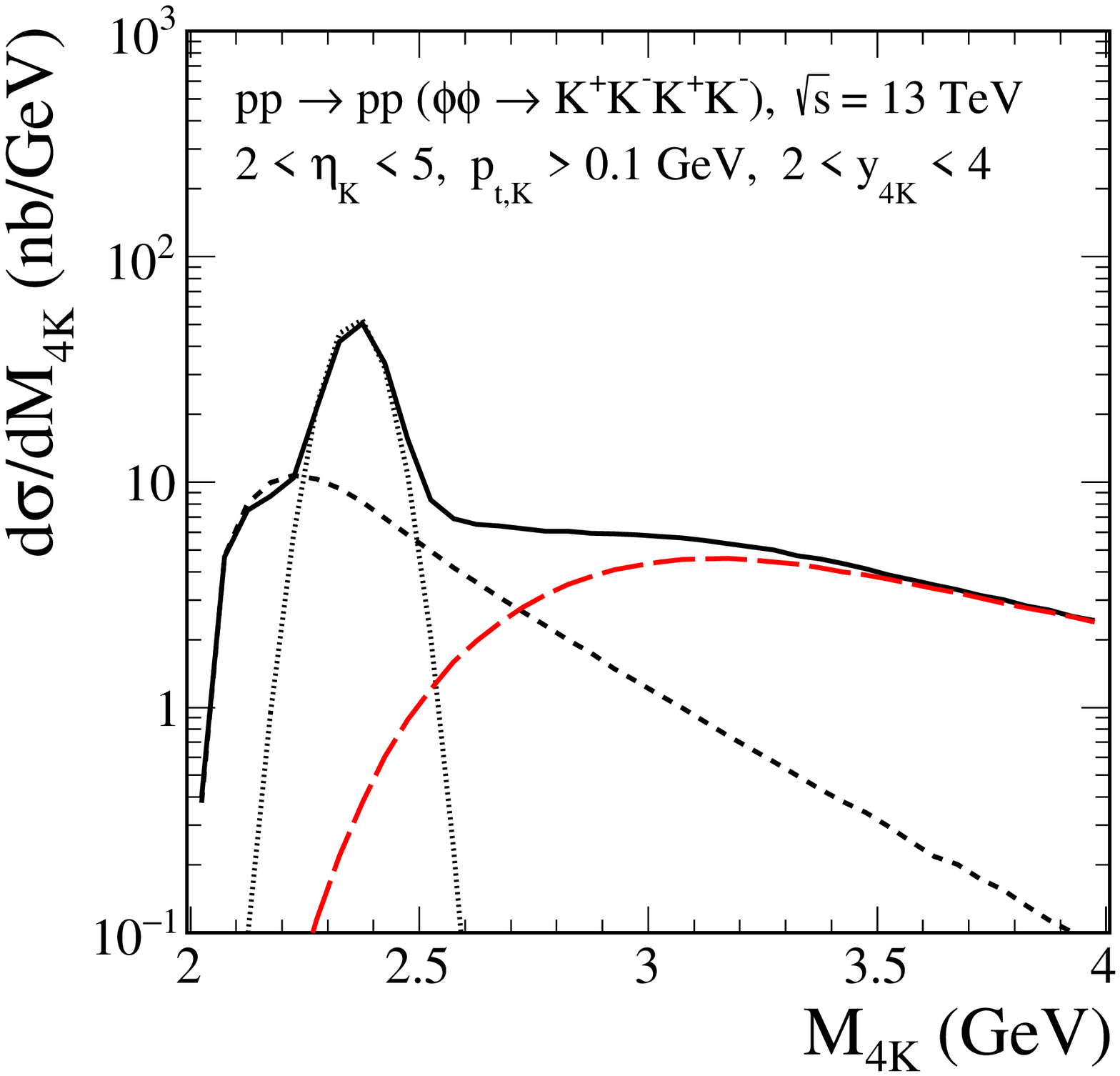}
\caption{The distributions in $\phi\phi$ invariant mass (left)
for $\sqrt{s} = 29.1$~GeV and $|x_{F,\phi \phi}| \leq 0.2$
and in ${\rm M}_{4K}$ (right) 
for $\sqrt{s} = 13$~TeV and $2 < \eta_{K} < 5$,
$2 < \rm{y}_{4K} < 4$, $p_{t, K} > 0.1$~GeV.
The WA102 experimental data from \cite{Barberis:1998bq} are shown.
The black long-dashed line corresponds to 
the reggeized $\phi$-exchange contribution 
and the black dashed line corresponds to the $f_{2}(2340)$ contribution.
The red dashed line represents the odderon-exchange contribution.
The coherent sum of all terms is shown by the black solid line.}
\label{fig:odderon_LHC}
\end{center}
\end{figure}

Now we go to the $pp \to pp K^{+}K^{-}K^{+}K^{-}$ reaction.
Fig.~\ref{fig:odderon_LHC} shows the results
including the $f_{2}(2340)$-resonance contribution
and the continuum processes due to reggeized-$\phi$ and odderon
exchanges.
For the details how to calculate these processes see \cite{Lebiedowicz:2019jru}.
Inclusion of the odderon exchange improves the description of the WA102 data \cite{Barberis:1998bq} for the $pp \to pp \phi \phi$ reaction;
see the left panel of Fig.~\ref{fig:odderon_LHC}.
In the right panel we show the distribution in four-kaon 
invariant mass for the LHCb experimental conditions.
The small intercept of the $\phi$-reggeon exchange, 
$\alpha_{\phi}(0) = 0.1$
makes the $\phi$-exchange contribution steeply falling 
with increasing ${\rm M}_{4K}$.
Therefore, an odderon with an intercept 
$\alpha_{\Ode}(0)$ around 1.0
should be clearly visible in the region 
of large ${\rm M}_{4K}$ 
(and also for large rapidity distance between the $\phi$ mesons).

\section{Conclusions}

We have discussed the possibility to search
for odderon exchange in the $p p \to p p \phi \phi$ and $p p \to p p \phi$ reactions
with the $\phi$ meson observed in the $K^+ K^-$ or $\mu^{+}\mu^{-}$ channels. 
For single $\phi$ CEP at the LHC there are two basic processes: 
the relatively well known (at the Born level) photon-pomeron fusion
and the rather elusive odderon-pomeron fusion.
In this context the photon-pomeron fusion is a background for 
the odderon-pomeron fusion.
The parameters of photoproduction
were fixed to describe the HERA $\phi$-meson photoproduction data; 
see \cite{Lebiedowicz:2019boz}.
There, we pay special attention 
to the importance of the $\phi$-$\omega$ mixing effect.

To fix the parameters of the pomeron-odderon-$\phi$ vertex
(coupling constants and cut-off parameters)
we have considered several subleading contributions
and compared our theoretical predictions for the $pp \to pp \phi$ reaction
with the WA102 experimental data from \cite{Kirk:2000ws}.
Having fixed the parameters of the model we have 
made estimates of the integrated cross sections as well as shown
several differential distributions at the LHC;
see Table~II of \cite{Lebiedowicz:2019boz}.
In our opinion several distributions should be studied 
to draw a definite conclusion on the odderon exchange.
For the $\phi \to K^+ K^-$ channel the distribution in $\rm{y_{diff}}$ 
(rapidity difference between kaons) seems particularly interesting.
It is a main result of our analysis that,
the $\rm{y_{diff}}$ distributions
are very different for the $\gamma$-$\Pom$- and $\Ode$-$\Pom$-fusion processes.
Observation of the pattern of maxima and minima 
would be interesting by itself as it is due to interference effects.
This should be a big asset for an odderon search.
The $\mu^+ \mu^-$ channel seems to be less promising
in identifying the odderon exchange 
at least when only the $p_{t, \mu}$ cuts are imposed.
To observe a sizeable deviation from photoproduction,
in the $\phi \to \mu^+ \mu^-$ channel,
a $p_{t,\mu^+ \mu^-} > 0.8$~GeV cut on the transverse momentum 
of the $\mu^+ \mu^-$ pair seems necessary. 
A combined analysis of both the $K^+ K^-$ and the $\mu^+ \mu^-$ channels
should be the ultimate goal in searches for odderon exchange.

The $pp \to pp \phi \phi$ process via odderon exchange
shown in Fig.~\ref{fig:4K_diagrams}~(a)
seems promising as here the odderon does not couple to protons.
In our analysis on two $\phi$-meson production 
in proton-proton collisions \cite{Lebiedowicz:2019jru} 
we tried to tentatively fix the parameters of 
the pomeron-odderon-$\phi$ vertex to describe the relatively large
$\phi \phi$ invariant mass distribution measured 
by the WA102 Collaboration \cite{Barberis:1998bq}.
Here we presented results for the odderon-exchange contribution
with fixed the parameters of our model to the WA102 data \cite{Kirk:2000ws}
on $pp \to pp \phi$; see Sec.~IV~A of \cite{Lebiedowicz:2019boz}.
We find from our model that the odderon-exchange contribution
should be distinguishable from other contributions
for relatively large rapidity separation between the $\phi$ mesons.
Hence, to study this type of mechanism one should investigate events with
rather large four-kaon invariant masses, outside of the region of resonances.
These events are then ``three-gap events'':
proton--gap--$\phi$--gap--$\phi$--gap--proton.
Experimentally, this should be a clear signature.

We are looking forward to first experimental 
results on single and double $\phi$ CEP at the LHC.

{\bf Acknowledgments}
I am indebted to Otto Nachtmann and Antoni Szczurek 
for collaboration on the issues presented here.
This work was supported by the 
Polish National Science Centre
grant 2018/31/B/ST2/03537.


\end{document}